# Gamma camera imaging in an undergraduate physics course

Mary Lowe, Alex Spiro and Peter Kutt

*Physics Department, Loyola University Maryland, Baltimore, MD 21210*

There are supplementary materials that will appear when this paper is published. The materials are not included with this preprint. If you wish more information, please contact mlowe@loyola.edu.

## Abstract

Gamma camera imaging is an important technique in nuclear medicine. It is capable of diagnostic imaging of metabolically active areas and organ function, and it can be used to evaluate blood flow in the heart muscle, measure bone growth, detect tumors, and perform many other medical studies. It is a real world application that integrates concepts in medicine, nuclear physics, geometric optics, data processing, calibration, and image formation. This paper provides an overview of gamma camera imaging intended for an intermediate-level undergraduate physics course for students majoring in STEM disciplines. Because working with radioactive materials is not practical in our setting, we use an approach involving paper-and-pencil exercises, a visible light apparatus, and computer work.

## I.   INTRODUCTION

Diagnostic nuclear medicine involves injection of drugs labeled with radionuclides into the human body. These drugs, called tracers or radiopharmaceuticals, are chosen to become localized in specific target tissues and provide information for a wide range of diseases. When the tracer emits gamma (γ) rays, they are detected by a gamma camera (also known as Anger camera), which produces a two-dimensional image of the three-dimensional spatial distribution of the tracer. The image reveals information about tissue function and biological processes underlying disease. Radionuclide imaging laboratories are now found in almost every hospital in the US.[1] More advanced gamma camera systems, known as single-photon emission computed tomography (SPECT), incorporate one or more rotating gamma camera heads to produce a set of planar images that are reconstructed into a three–dimensional image. The most common uses of SPECT are static and dynamic imaging studies[1] to diagnose or monitor disorders of the brain, heart, bone, thyroid, lung, liver, and abdomen.[2] Gamma camera imaging can obtain quantitative information about physiological function and the molecular basis of disease without perturbing processes within the body.[3] Other diagnostic techniques, such as X-rays, computed tomography, and magnetic resonance imaging (MRI), show anatomical changes due to an abnormality.

Gamma camera imaging (GCI) is an important application of nuclear physics and a major topic in the medical physics curriculum. Typically, however, it is not included in the undergraduate physics curriculum. This paper describes how we have introduced GCI in an intermediate-level physics course appropriate for undergraduates majoring in science and engineering disciplines who have had 1–3.5 years of physics. By including more applications of physics in the undergraduate curriculum, we hope to help students majoring in other disciplines appreciate the importance and relevance of physics.

Our semester-long course surveys diagnostic and therapeutic techniques used in medicine, including eyeglasses, fiber optics, ultrasound imaging, computed tomography (CT), positron emission tomography (PET), MRI, in addition to GCI. The portion devoted to nuclear physics, PET and GCI lasts about four weeks. During that time, we cover many of the topics found in the nuclear physics chapter of a standard introductory physics textbook. For the portion on radioactive decay, for example, we relate the half-lives of radionuclides to medical applications, [14]C-dating, nuclear power, and nuclear weapons. The penetration depth of $\alpha$, $\beta$, and $\gamma$ radiation through various materials is also discussed.

For most undergraduate programs, the cost of a real gamma camera is prohibitive, and it is not feasible for a large class of untrained students to work with radioactive materials. We therefore designed several types of apparatus that use visible light to model the key physical principles of the gamma camera. For GCI, the level of presentation in Kane[4] is appropriate for our students, but we have enriched the reading with experimentation, demonstrations, paper-and-pencil and computer activities, and lectures. Other useful references are Cherry et al[1] and Lodge and Frey.[5]

## II.    OVERVIEW

### A. Radiopharmaceuticals

To perform GCI, a drug containing a $\gamma$-ray-emitting radionuclide is introduced into the body. Radiopharmaceuticals are an active area of research and are designed to be preferentially absorbed by certain organ systems and disease pathways. For most applications, the $\gamma$-ray photon has an energy in the 30 keV – 250 keV range.[6] For example, one of the primary radionuclides used with a gamma camera is technetium-99m, which emits 140 keV $\gamma$ rays and has a half-life of 6.02 hrs.[1]

### B. Gamma camera

The simplest gamma camera consists of one head, as shown in Fig. 1. The components are housed in a box that is positioned on one side of the patient's body. The tracer inside the patient emits $\gamma$ rays isotropically. A lead or tungsten collimator, consisting of an array of holes, constrains the angle of $\gamma$ rays that can pass upwards into the detection system. The collimator is essential for producing images of the spatial distribution of tracer particles. After passing through the collimator, gamma radiation enters a large scintillator typically made of a NaI(Tl) crystal. When absorbed in NaI(Tl), each $\gamma$-ray photon creates a burst of blue fluorescent photons. The blue photons pass through a light guide and are detected by 30–90 photomultiplier tubes (PMTs) arranged in a two-dimensional (2D) array, which convert the scintillator light into electrical pulses.[5] The light guide enables the blue burst to be spread across multiple PMTs, which turns out to be critical for good spatial resolution (see Section III. B. 3). The detector electronics shapes the pulses, digitizes the pulse heights, and sends the digital signals to a computer where the x-y position of each $\gamma$-ray photon is calculated. The coordinates of all photons are binned into a 2D array of pixels and a histogram is constructed.[1] The net result is an image of the body where the pixel intensity is proportional to the number of $\gamma$-ray photons in each pixel bin (Fig. 2). 10,000-20,000 counts are recorded each second, depending on the study. The challenge is to teach students how PMTs with a diameter of several centimeters can image the tracer distribution with a spatial resolution of a few millimeters.

## III.     TEACHING GAMMA CAMERA IMAGING TO UNDERGRADUATES

We divide the curriculum into three sections: (1) collimation, (2) detection, and (3) computer processing. We model the gamma camera with visible light replicas, allowing students to see directly how the rays propagate through the apparatus. The components used in the apparatus are listed in the supplementary materials.[7]

### A. Collimator and the point spread function

After the radiopharmaceutical is administered to the patient, it accumulates in certain regions of the body and emits $\gamma$ rays. The goal is to measure the locations of the tracer. Even though $\gamma$ rays are electromagnetic radiation, a γ-ray image cannot be created using traditional optical techniques because gamma emission cannot be focused, and the direction of γ rays is difficult to determine.

The collimator is introduced using a paper-and-pencil exercise for a one-dimensional (1D) array of holes and a single point source S that emits γ rays isotropically (Fig. 3). Students do a ray tracing exercise to determine whether γ emission passes through the collimator or is blocked by its lead septa, which results in a narrowly directed propagation of γ-ray photons at the output of the collimator. The number of photons passing through the $i^{th}$ hole ($i = 0, \pm1, \pm2, \dots$) is proportional to $\theta_i$, the angle subtended by that hole. Students measure $\theta_i$ with a protractor, construct a bar graph of the distribution of γ rays and plot each value at the center of the hole. To match drawings in textbooks and learn about fitting, they then fit the data to a Gaussian function (Fig. 3) and estimate the full width (FW) of the point spread function (PSF). The Gaussian is used in a nuclear medicine physics handbook.[5] Other ray tracing exercises, with varying levels of difficulty, can also be assigned exploring the dependence of the PSF on geometric parameters $d$, $b$, and $h$.

While the true PSF must be calculated in three dimensions, we introduce it by deriving the irradiance function $I(x)$ for the planar collimator geometry shown in Fig. 3, in which the collimator holes are spaced by period $p$ along the x-direction, and the analysis is confined to a plane. This approximation shows the essential features of the problem and is easier to visualize. The cross-sectional shape of real collimator holes does not enter this analysis. The derivation appears in the supplementary material.[8] The result is:

$$I(x) = I_o \, tan^{-1}\left[ \frac{dL - |x|h}{(L^2 + x^2) - \frac{1}{4}(h^2 + d^2)} \right], \quad \frac{d}{2} \leq |x| \leq \frac{dL}{h}$$

$$I(x) = I_o \, tan^{-1}\left[ \frac{d\left(L + \frac{h}{2}\right)}{\left(L + \frac{h}{2}\right)^2 + \left(x^2 - \frac{1}{4}d^2\right)} \right], \quad |x| \leq \frac{d}{2}$$

(1)

Source S can be located anywhere beneath the collimator. The x-axis is chosen so that $x = 0$ is located at S. $I(x_i)$ is the irradiance at $x_i$, where $x_i$ is the position of the center of the $i^{th}$ hole. $b$ is the vertical distance between the source and the bottom of the collimator; $d$ is the hole width; h is the thickness of the collimator; $L = b + h/2$; and $I_o$ is fitted to the data. The angle subtended by each hole $i$ is equal to $I(x_i)/I_o$. For the geometry shown in Fig. 3, $x$ equals 0 for the hole directly above S, and $x = \pm p$, $\pm 2p$ and $\pm 3p$ for the adjacent holes. The quantity in the square brackets of Eq. (1) corresponds to $tan\ \theta_0$, $tan\ \theta_1$, etc. The solid dots in the figure represent the irradiance, given by Eq. (1), at each hole.

The FW of the PSF can be determined approximately by inspecting the geometry shown in Fig. 3. Angle $\alpha$ is the upper limit for radiation to pass through the collimator. Therefore,

$$\tan\alpha = \frac{d}{h} = \frac{\frac{1}{2}FW_{PSF}}{b+h}$$

or                                                                                                      (2)

$$FW_{PSF} = 2\frac{d(b+h)}{h}$$

A real gamma camera collimator consists of thousands of holes with round, hexagonal or square cross sections.[5] Typical dimensions are: hole diameters 0.2 mm to 3 mm; hole lengths 30 mm to 58 mm, and septal thickness 0.1 mm to 1.05 mm.[9,10] Our model of a 2D collimator consists of a hexagonal array of round holes (Fig. 4a) created by drilling a dark gray plastic sheet (3/16" thick) with a laser. The number of stacked plastic sheets can be varied. Tracer point sources are represented by three LEDs (Vernier Color Mixer kit, Fig. 4b) with a 9.2 mm center-to-center distance. The brightness of the LEDs can be adjusted independently. A sheet of paper taped to the collimator output shows the hexagonal pattern of the rays that pass through the collimator (Fig. 5a).

Suppose the goal is to image the heart, and there are two tracer point sources separated by a distance $tr$ along the x-axis. Each source is associated with a PSF. What is the minimum $D_{source}$ such that the two PSFs can yield medically useful information? To address this issue of spatial resolution using the model apparatus, students use two LEDs and vary $b$ or $h$. In Fig. 5b top ($b \approx 5$ cm), the PSF spots are well separated. In Fig. 5b bottom ($b \approx 10$ cm), the $FW_{PSF}$ is broader and the spots start to merge. By sketching qualitative graphs of the intensity profiles of the two spots, students attempt to find a criterion for when the spots are barely distinguishable. The conclusion is related to the Rayleigh criterion discussed in introductory textbooks.[11] The collimator resolution[1] is approximately ½ $FW_{PSF}$, which improves when $b$ or $d/h$ decreases (see Eq. (2)). A demonstration of an L-shaped source is shown in Fig. 5c, leading to a discussion of medical applications of imaging.

For a single LED, the $FW_{PSF}$ can be experimentally determined by measuring the distance between the outermost illuminated holes that appear on the paper sheet of the collimator along the center horizontal line of the spot, as shown by the arrow in Fig. 5a. As $b$ is increased, the criterion for measuring the $FW_{PSF}$ is the appearance of rays at the far edges of the outermost holes (Fig. 6); the distance is equal to *(n-1)p + d*, where n is the number of holes. A comparison between experimental results and the prediction of Eq. (2) is shown in Fig. 6. The accuracy is limited by the finite size of the light source and by the uncertainty in the source-to-collimator distance $b$.

The collimator can serve as a rich source of challenge problems. (a) For the planar geometry shown in Fig. 3, what is the irradiance at the output of each hole as the source is moved parallel to the array of holes? The answer is given in the supplementary materials.[8] (b) Consider a 3D collimator with a 1D array of holes with a fixed depth and a rectangular or circular cross section, illuminated by an isotropic point source. How does the length of the holes *transverse* to the direction of the array affect the irradiance? In the supplementary materials,[8] we derive the exact irradiance function for rectangular holes and fit it to the experimental data of Fig. 6. The results for square and circular holes are similar but the irradiance for circular holes drops off a little faster with $x$. (c) A Python program template for simulating light patterns from planar apertures is available in Ref. 12. Extending this template, simulate the light pattern from a 3D collimator with cylindrical holes.  See the supplementary materials[13] for sample assignments that have been completed by undergraduates.

**B. Detection system and determination of xy coordinates of a γ-ray photon**

After a γ-ray photon passes through the collimator, it is detected by a system comprised of a scintillator crystal, a light guide, an array of photomultiplier tubes (PMTs), electronics, and computer. We developed a model apparatus (Fig. 7) for students to visually see how detection occurs and to understand how to compute the location of the tracer to within a fraction of the diameter of a PMT.

***1. Fluorescent red plastic as analogue of scintillator crystal***

Typically a scintillator consists of a single crystal of sodium iodide doped with thallium, with a thickness of 0.95–1.5 cm and a lateral dimension ranging from the size of a thyroid ($\approx$ 10 cm) to the width of a human body (40–60 cm). When a γ-ray photon interacts with the scintillator crystal, free electrons are produced that excite electrons from the crystal's valence band to the conduction band. The dopant thallium creates additional energy levels within the band gap. An electron in the conduction band can fall down to the valence band via these intermediate levels, causing the emission of a photon with wavelength in the 350-500 nm range.[14] For each γ-ray photon that interacts with the scintillator crystal, thousands of violet-blue photons are produced isotropically. The index of refraction of NaI is 1.839 at 435.8 nm.[15]

We model this process (Fig. 7) using a green diode laser (λ= 532 nm, 3-5 mW) aimed upwards to mimic γ emission that has passed through a collimator hole. A sheet of red fluorescent acrylic (Estreetplastics, 0.230" thickness, λ = 599 nm at fluorescence maximum) acts as the scintillator crystal and emits orange photons isotropically when excited with green laser photons. Students will notice that fluorescence occurs at the lower boundary of the material, but the instructor needs to point out that in an actual scintillator crystal, the emission of visible light can occur anywhere within its thickness.

## 2. Acrylic sheet as analogue of light guide

We use an acrylic rod with a 2" diameter to mimic a PMT; this diameter is approximately the same as that of PMTs in a real gamma camera. Students will believe that to detect the photons, PMTs should be placed directly on the scintillator. A quick test with the model apparatus will show that imaging cannot be done in this way. When an array of PMT rods is placed directly on top of the fluorescent plastic, essentially all of the light exiting the plastic is collected by the rod directly over the laser due to total internal reflection (TIR) within the plastic, and the rays exiting the scintillator crystal are concentrated above the fluorescence (Fig. 8a). Even if the effects due to TIR are reduced with a layer of water in between the plastic and the PMT rods, one rod still collects most of the light due to the small distance between the fluorescence and the rod. Therefore the location of the laser ($\gamma$ ray) would only be determined to within a 2" diameter circle. This is inadequate for most medical purposes.

For medical imaging, the location of each $\gamma$-ray point source must be determined to within a few millimeters. A light guide is used to spread the scintillator light across the array of PMTs. As shown in a ray tracing exercise, by having a light guide and optical grease between the scintillator and light guide, and between the light guide and PMTs (Fig. 8b), TIR is eliminated. If the light guide is thick enough, the light will spread across multiple PMTs.

In the model apparatus (Fig. 7), the light guide is a commercially available, 1.302" thick, clear acrylic sheet ($n_{clear}$ = 1.46 $\pm$ 0.02 at 650 nm, McMaster-Carr). A few milliliters of water are placed on the fluorescent plastic ($n_{fluor}$ = 1.48 $\pm$ .02 at 650 nm, $\theta_c$ = 42.4° ), and the light guide is positioned on top in such a way that the water can spread uniformly with no bubbles. The water mimics optical grease, reducing TIR by replacing air at the interface of the two sheets. The thickness of the clear acrylic must be sufficient for there to be a measurable amount of light over multiple PMT rods.

### 3. Centroid algorithm and 1D array of 2" acrylic rods as analogue of 1D PMT array

To learn how to determine the location of a γ-ray photon, students use a model apparatus consisting of a 1D array of five 2"-diameter acrylic rods, each with a length of 4" and unpolished end faces, to mimic a 1D PMT array (Fig. 7c). Measurements are improved by placing a moist paper towel between the light guide and rods to diffusively scatter the light and spread the light among more rods. Without the paper towel, we believe there are microscopic air bubbles in the water layer at the end face that lead to TIR. The paper towel also reduces dripping as the students work with the apparatus, and markings on the paper towel help align the rods and laser. As the laser is moved, mimicking different positions of the γ-ray photon, variations in light intensity are seen at the top of the PMT rod array, and students can intuitively understand where the light source is located.

To calculate the position of the light source quantitatively, a silicon photodiode detector (Thorlabs DET100A) is used to measure the light intensity at the top of each PMT rod. A reflective cone is mounted at the photodiode input to collect light from the entire end face of the rod (Fig. 7a). To determine the position of the light source, a centroid calculation is performed using $N$ voltage values obtained from $N$ PMT rods ($N$ = 5 in Fig. 7c):

$$\overline{x} = \sum_{i=1}^{N} s_i x_i \Big/ \sum_{i=1}^{N} s_i \qquad (3)$$

where $s_i$ = signal from the $i^{th}$ PMT (with background subtracted) centered at $x_i$.

Fig. 9 shows that the calculated x positions are consistently less than the true values, with the discrepancy increasing for larger x. This is due to the large diameter of the PMT rods and the small number of PMT rods in the teaching apparatus. Multiple types of nonlinearities exist in gamma cameras, and various procedures and algorithms are used to improve position accuracy.[1] Knoll obtained a graph similar to Fig. 9 for a real gamma camera, using a cubic spline for interpolation.[16] We developed a simpler procedure that uses the polynomial fitting feature in Excel; students are given calibration data to determine the degree of the polynomial that best fits the data. Then they position the laser at an arbitrary location, acquire voltages from the PMT rods, do a centroid calculation, and use the polynomial fit to determine a more accurate location of the laser. The result is sensitive to the alignment of the laser and PMT rods. While a 5[th] degree polynomial is adequate for the graph in Fig. 9, we do not expect a polynomial will fit the data well for a larger array of rods.

### 4. Centroid algorithm and 2D array of 2" acrylic rods as analogue of 2D PMT array

A real gamma camera incorporates a hexagonal array of 30-90 PMTs. We model this using 19 PMT rods arranged in a hexagonal lattice (Fig. 7a). As the laser source is shifted manually, different PMT rods are illuminated (Fig. 7b). Calculation of the source position ($\overline{x}, \overline{y}$) is achieved with a 2D centroid calculation that is equivalent to the algorithm originally developed by Anger. See supplementary material[17] for a proof. In our classes, we demonstrate the 2D gamma camera, teach students how to calculate the centroid, and provide them with a dataset of light intensity values from 19 rods. Then for homework, they calculate the position of the laser using Excel. Our students do not correct for distortion but this topic could be explored in an advanced lab.

### C. Gamma camera electronics

Gamma camera electronics consists of photon counting detectors operating in pulse mode, where each event, i.e., each interaction between a $\gamma$-ray photon and the scintillator crystal, is processed individually. There are multiple levels of detail for presenting the electronics used for detection of $\gamma$-ray-emitting radionuclides, and instructors must decide which level is suitable for their students. A discussion of pedagogical aspects appears in Parks and Cheney.[18] A brief overview is given here.

The energy of a $\gamma$-ray photon is deposited in the scintillator crystal, emitting optical photons isotropically in a quantity proportional to the absorbed energy. The optical scintillation photons are detected by the PMT array, where the output of each PMT is a current pulse with a duration of a few nanoseconds and a total charge that is proportional to the number of scintillation photons striking the tube's photocathode.[14] Since the scintillation photons from each event are detected by multiple PMTs in the array, the sum of the outputs from all PMTs is proportional to the total energy deposited by the $\gamma$-ray photon.[4] Pulse-height-analysis of the summed outputs reveals a photopeak occurring when the full energy of the $\gamma$-ray photon is converted to visible light within the scintillator. However, the response function of the scintillator is more complex and includes effects such as Compton scattering within the patient or the scintillation crystal, X-rays from impurities or lead shielding, backscatter, and object scatter within the patient.[1,14] To improve the signal-to-noise ratio and contrast in the gamma camera image, a discriminator (i.e., window) is set around the photopeak. For each event within the photopeak window, the voltage pulses from the PMTs of the array are processed collectively to determine the position of the $\gamma$-ray photon.[1] The resolution of the final image is determined by the intrinsic resolution of the detection system and the resolution of the collimator; the latter dominates the total spatial resolution.[5]

As an example, each radioactive decay of $^{99m}$Tc produces 140 keV of emitted $\gamma$-ray energy. The scintillator NaI(Tl) produces approximately $4.7 \times 10^3$ visible photons per $\gamma$-ray photon, or one visible photon per 30 eV of $\gamma$-ray energy.[19] The discriminator window is set to $140 \pm 10$ keV in order to reject noise and pulses resulting from scatter.[1] Event rates on the order of $10^5$ $\gamma$-ray photons per second[6] can be processed by the gamma camera electronics.

**D. Computer simulation of a gamma camera scan with moving bed**

To help students gain a better understanding of how an image is formed using a gamma camera, a simulation of a patient with pheochromocytoma, a tumor of the adrenal glands, is included in the supplementary materials.[20] Simulation "wholebodybed.py" was written in Python 3.7, VPython and associated libraries obtained from the Anaconda3 distribution. The user can edit the count rate, number of points, etc. Binning into a pixel array is not done in the simulation. A movie file of the simulation is included in the supplementary material along with a description.[20] The simulation, which is based on the work of Buck et al,[21] shows a simplified version of a whole body gamma camera scan in which the position of each $\gamma$-ray photon is plotted on the screen at a rate slow enough for students to see how the image is created dot-by-dot over time, where each dot corresponds to the position of a $\gamma$-ray photon that passes through a collimator hole. The patient is initially positioned so that the head and neck are imaged. Gradually, the bed moves so that the torso followed by the legs are scanned, and a planar view of the entire body forms over time. The gamma camera scan shows a high concentration of a radiopharmaceutical on the right side of the torso (left side in the image) in the vicinity of the kidney and bladder.

While not clearly visible in the simulation, Buck et al state that the gamma camera and SPECT images show two lesions in the adrenal glands. They also show the importance of multimodal imaging techniques such as SPECT/CT, in which functional and anatomical images are co-registered to locate the disorder more accurately.

## IV.    NEW DIRECTIONS

GCI is an evolving technique. Since its inception, there have been several important innovations. The positioning of the gamma camera over the patient and the operation of the equipment have become more automated. The need for three-dimensional information has led to the growth of SPECT, in which multiple gamma camera scans are conducted from different angles around the body. Algorithms, known as filtered back projection (FBP),[1] are applied to a set of planar scans to reconstruct images of thin cross sections of the body. Fig. 10 shows an example of how SPECT can image blood flowing through heart muscle using a procedure known as myocardial perfusion imaging (MPI).[22] On the left, the plane of the image slice is shown, which cannot be acquired by a single gamma camera scan; in the reconstructed image slice on the right, the wall of a healthy left ventricle appears doughnut-shaped. In our course, we discuss SPECT because of its prevalence in hospitals and in the literature. Filtering refers to mathematical operations applied to the images to reduce blurring, smooth out noise, etc. While concepts underlying FBP are outside the scope of this paper, we cover the basic ideas in class, in the unit on CT, and also assign readings in Kane.[4] We demonstrate the effects of filtering using software.

New detector technologies are being developed that are capable of MPI, small-animal imaging, molecular breast imaging, and other applications that require high spatial resolution, good image quality in a practical length of time, and lower doses of radiopharmaceuticals.[6] Solid-state detectors for gamma cameras have been introduced, in which $\gamma$-ray photons are absorbed by a semiconductor material such as cadmium zinc telluride (CZT). The CZT is pixelated (pixel size 2.5 x 2.5 mm) with a direct readout from each pixel. A collimator is still required but no scintillation crystal, PMTs, or centroid calculation are needed.[22]

There are also specialized gamma cameras dedicated to cardiology studies. For example, the D-SPECT camera uses square-shaped detectors arranged in a semicircle around the heart. The detectors are composed of nine CZT blocks that move independently.[22] This is advantageous because in a gamma camera designed for whole body scanning, only a small fraction of the collimator and scintillator are used to obtain the heart image.[23] A camera specifically designed for the heart can be situated closer to the patient's chest and allows more flexibility in patient positioning,[22] resulting in superior images.

**ACKNOWLEDGMENTS**

We wish to thank Martin Lodge and Aimei Kutt, and also acknowledge Yanko Kranov, Nancy Donaldson, Randy Jones, Patrick Doty, Jay Wang, and Ernie Behringer for their many contributions. This project was supported by NSF TUES award DUE-1140406.

**FIGURE CAPTIONS**

Figure 1. Basic components of the gamma camera.

Figure 2. Upper body, planar bone scan using the radiopharmaceutical $^{99m}$Tc-methylene diphosphonate. In this posterior view, over- and underlying organs are superimposed on each other. The image is constructed photon-by-photon. The original bone scan of the whole body is located in the supplementary material.[24]

Figure 3. Exercise to determine the point spread function of a γ-ray source. Students are given a 1D collimator and a point source S on a piece of paper. The dark shaded regions are lead septa opaque to γ rays. The number of γ-ray photons passing through hole $i$ is proportional to the angle $\theta_i$ subtended by each collimator hole. Students construct a bar graph of the intensity (stripe) and fit a Gaussian to the measured intensities (solid line). The solid dots correspond to the exact irradiance function values for each hole (Eq. (1)). Equation (2) for FW$_{PSF}$ is determined from this geometry.

Figure 4. Parallel hole collimator. (a) The holes are laser-drilled into a 2" x 2" x 3/16" gray PVC Type 1 plastic to form a hexagonal lattice. The sheets are stacked and aligned with drill rods, and the assembly is held together with tape or rubber bands. The hole diameter is 1.2 mm; the nearest-neighbor, center-to-center spacing $p$ is 2.60 mm; the row separation is 2.25 mm. (b) Apparatus for viewing the output of the collimator. Light source A consists of red, green, and blue LEDs, mounted on an optical bench. The center-to-center distance $D_{source}$ is 9.2 mm. Collimator B is situated $b$ = 3 to 20 cm away from A. A sheet of white paper C is taped onto B at the output. Students view C.

Figure 5. Output of the collimator for small and large values of b. Distance between holes $p$ = 2.60 mm. (a) Single LED source, $d$ =1.2 mm, $h$ =18.3 mm, $b \approx 4$ cm (top), $b \approx 10$ cm (bottom). Each arrow marks the center horizontal line of the spot. (b) Two LEDs, with $D_{source}$ = 9.2 mm, showing the effect of the PSF upon the ability to distinguish the sources at the output. $d$ =1.2 mm, $h$ = 18.3 mm, $b \approx 5$ cm (top), $b \approx 10$ cm (bottom). (c) Five LEDs arranged in an L-shape. As b increases, the L becomes less clear.

Figure 6. Measurements of $FW_{PSF}$ for varying $b$. Hole diameter $d$ = 2.1 mm; thickness of collimator $h$ = 75.1 mm. The photo shows rays at the far edges of the outermost holes of the illuminated area at the output of the collimator. The FW is measured from the left edge to the right edge.

Figure 7. Detection system for the teaching apparatus. (a) Two-dimensional model. A green laser, representing a $\gamma$-ray photon that has passed through the collimator, causes orange fluorescence that passes through a light guide into a hexagonal lattice of cylindrical rods mimicking a PMT array. The rods are coated in silver paint or aluminum foil. (b) Output of 19 "PMT" rods. (c) One-dimensional model with five PMT rods. Water mimics optical grease and is located between the fluorescent plastic and the light guide. A wet paper towel is placed between the light guide and rods to scatter the light more uniformly. Graph paper is placed on the breadboard to align the laser and PMT rods and to measure the position of the laser.

Figure 8. Ray tracings done by students to show the purpose of the light guide. S represents the burst of visible photons produced by one $\gamma$-ray photon. For the model apparatus, fluorescence occurs at the bottom of the plastic, but in a scintillator crystal, the photon burst may occur anywhere in the depth of the material. (a) No light guide and no optical grease. TIR is present. (b) Light guide and optical grease cause the rays to spread across multiple PMTs.

Figure 9. importance of corrections in GCI. The calculated position of the light source is compared to the known position for a 1D array of PMT rods. The source is shifted right starting from the center of the apparatus at $x = 0$. Several positions are shown in the inset. The solid line is the ideal case where the calculated position is equal to the location of the light source. The dotted line is a $5^{th}$ degree polynomial fit of the true positions vs. the measured positions.

Figure 10. Cross-sectional image of the heart using SPECT. Image of the left ventricle wall was obtained using MPI, a technique that is conducted to determine areas of damaged coronary muscle. Image plane (left) and light gray regions (right) show where blood is flowing in the muscle. A full set of color image slices from three perpendicular directions appears in the supplementary material.[24]

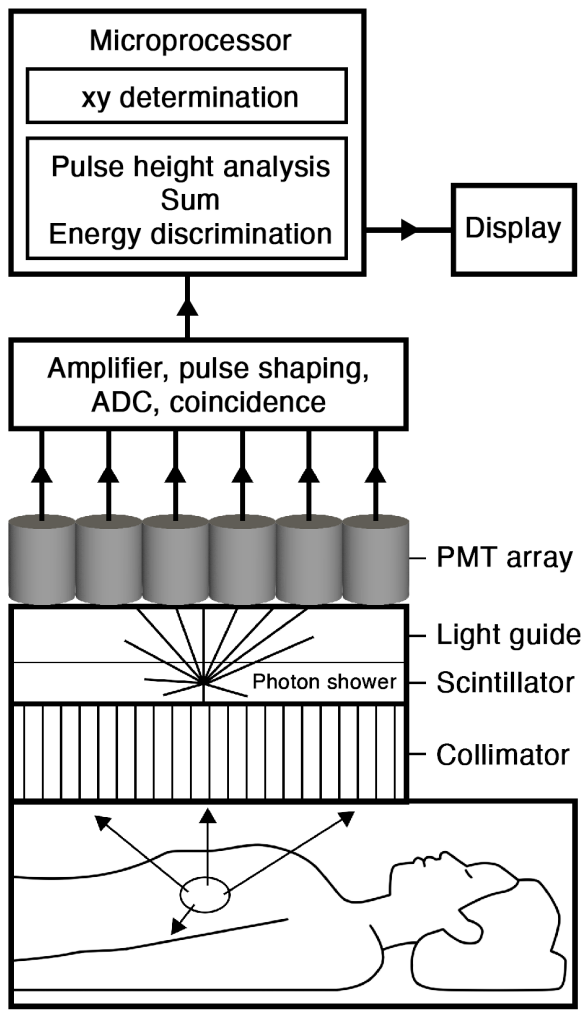

Figure 1. Basic components of the gamma camera.

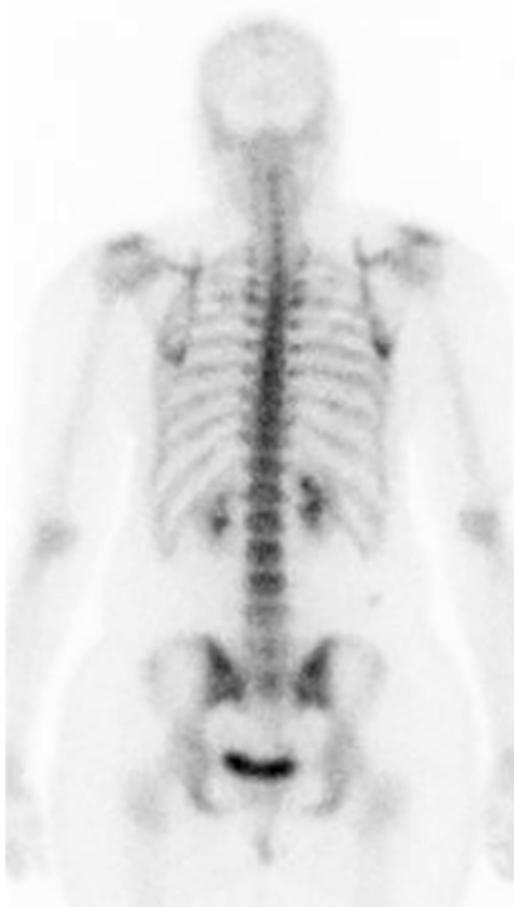

Figure 2. Upper body, planar bone scan using the radiopharmaceutical [99m]Tc-methylene diphosphonate. In this posterior view, over- and underlying organs are superimposed on each other. The image is constructed photon-by-photon. The original bone scan of the whole body is located in the supplementary material.[24]

Figure 3. Exercise to determine the point spread function of a γ-ray source. Students are given a 1D collimator and a point source S on a piece of paper. The dark shaded regions are lead septa opaque to γ rays. The number of γ-ray photons passing through hole i is proportional to angle $\theta_i$ subtended by each collimator hole. Students construct a bar graph of the intensity (stripe) and fit a Gaussian to the measured intensities (solid line). The solid dots correspond to the exact irradiance function values for each hole (Eq. (1)). Equation (2) for $FW_{PSF}$ is determined from this geometry.

(a)

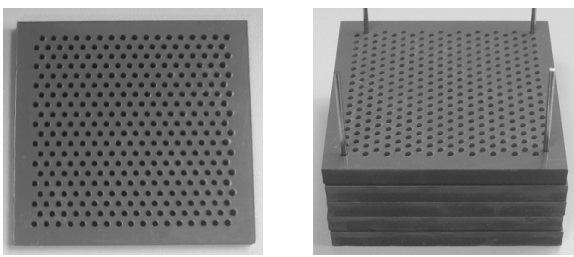

(b)

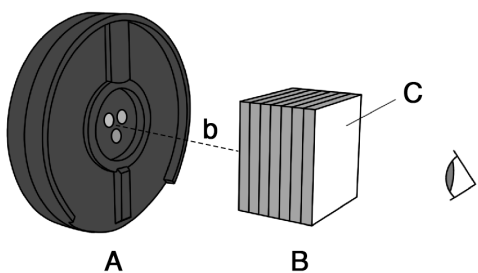

Figure 4. Parallel hole collimator. (a) The holes are laser-drilled into a 2" x 2" x 3/16" gray PVC Type 1 plastic to form a hexagonal lattice. The sheets are stacked and aligned with drill rods, and the assembly is held together with tape or rubber bands. The hole diameter is 1.2 mm; the nearest-neighbor, center-to-center spacing $p$ is 2.60 mm; the row separation is 2.25 mm. (b) Apparatus for viewing the output of the collimator. Light source A consists of red, green, and blue LEDs, mounted on an optical bench. The center-to-center distance $D_{source}$ is 9.2 mm. Collimator B is situated $b$ = 3 to 20 cm away from A. A sheet of white paper C is taped onto B at the output. Students view C.

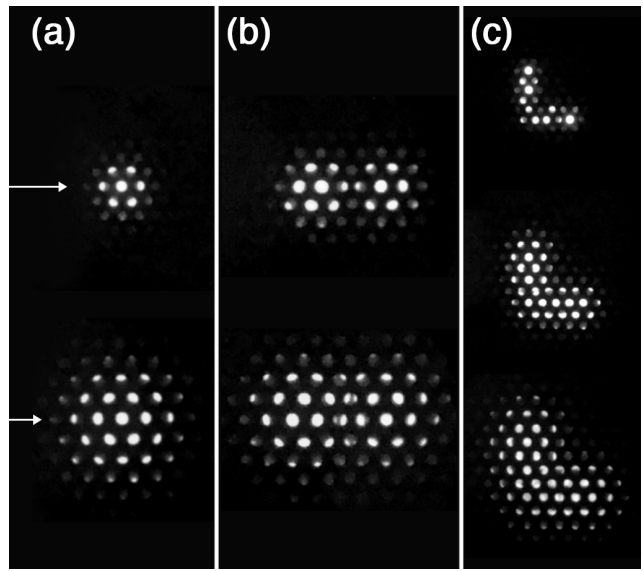

Figure 5. Output of the collimator for small and large values of b. Distance between holes $p$ = 2.60 mm. (a) Single LED source, $d$ =1.2 mm, $h$ =18.3 mm, $b \approx 4$ cm (top), $b \approx 10$ cm (bottom). Each arrow marks the center horizontal line of the spot. (b) Two LEDs, with $D_{source}$ = 9.2 mm, showing the effect of the PSF upon the ability to distinguish the sources at the output. $d$ =1.2 mm, $h$ = 18.3 mm, $b \approx 5$ cm (top), $b \approx 10$ cm (bottom). (c) Five LEDs arranged in an L-shape. As $b$ increases, the L becomes less clear.

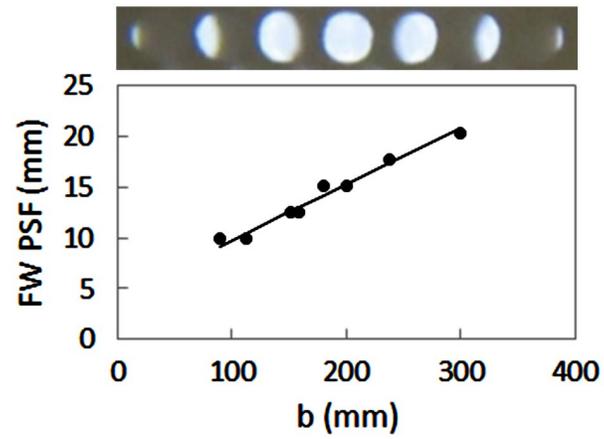

Figure 6. Measurements of FW$_{PSF}$ for varying $b$. Hole diameter $d$ = 2.1 mm; thickness of collimator $h$ = 75.1 mm. The photo shows rays at the far edges of the outermost holes of the illuminated area at the output of the collimator. The FW is measured from the left edge to the right edge.

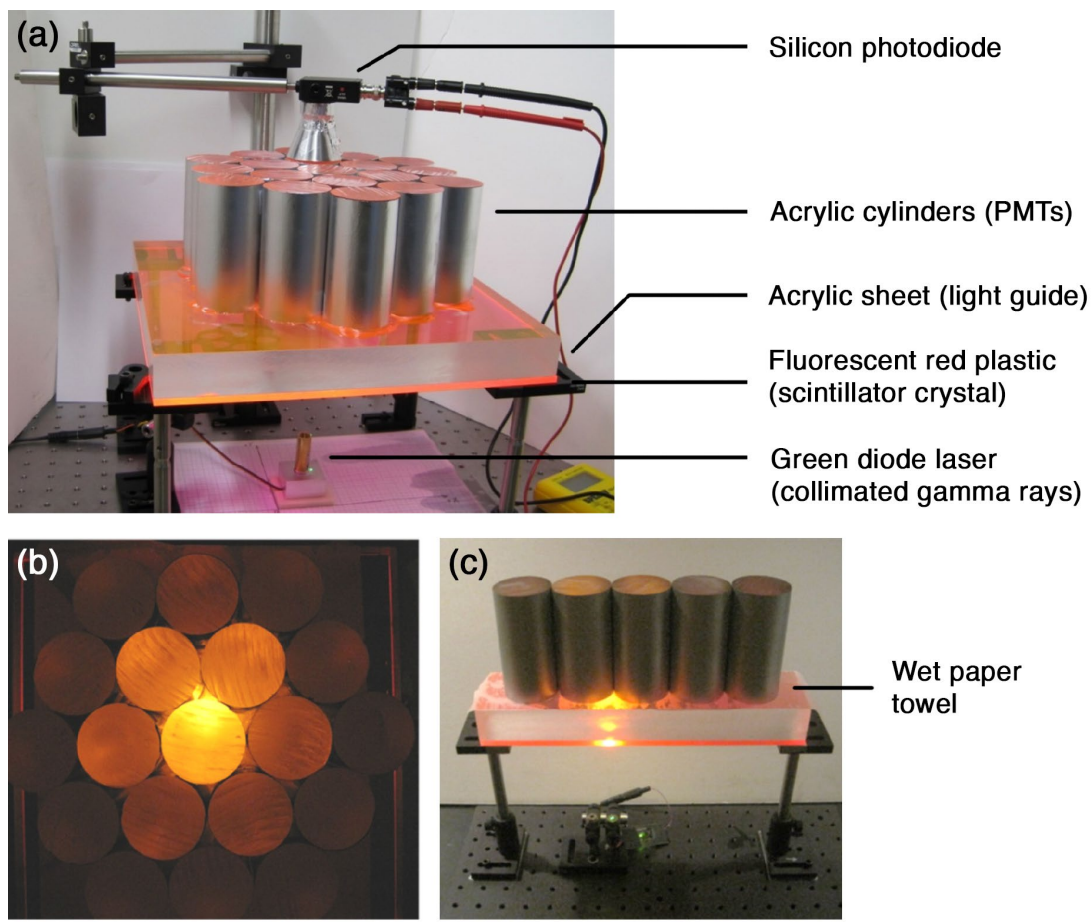

Figure 7. Detection system for the teaching apparatus. (a) Two-dimensional model. A green laser, representing a γ-ray photon that has passed through the collimator, causes orange fluorescence that passes through a light guide into a hexagonal lattice of cylindrical rods mimicking a PMT array. The rods are coated in silver paint or aluminum foil. (b) Output of 19 "PMT" rods. (c) One-dimensional model with five PMT rods. Water mimics optical grease and is located between the fluorescent plastic and the light guide. A wet paper towel is placed between the light guide and rods to scatter the light more uniformly. Graph paper is placed on the breadboard to align the laser and PMT rods and to measure the position of the laser.

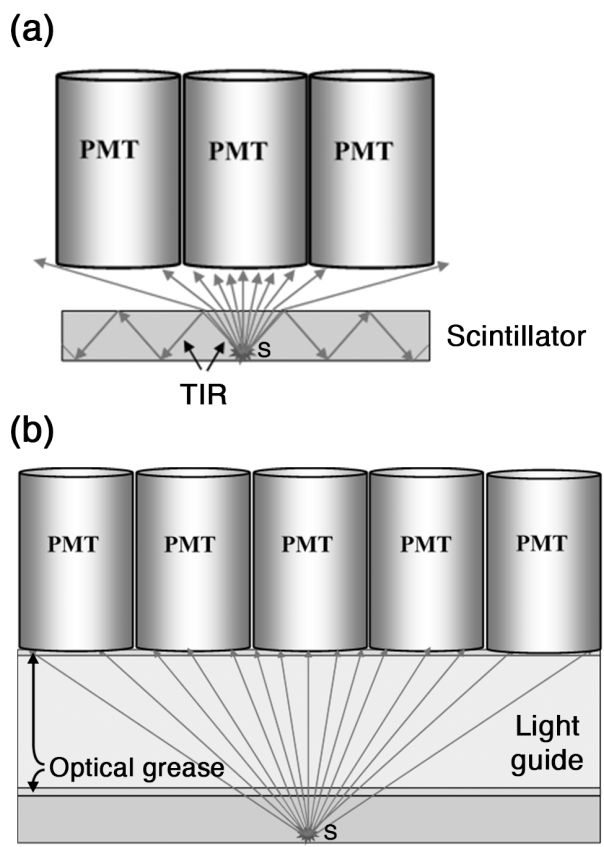

Figure 8. Ray tracings done by students to show the purpose of the light guide. S represents the burst of visible photons produced by one γ-ray photon. For the model apparatus, fluorescence occurs at the bottom of the plastic, but in a scintillator crystal, the photon burst may occur anywhere in the depth of the material. (a) No light guide and no optical grease. TIR is present. (b) Light guide and optical grease cause the rays to spread across multiple PMTs.

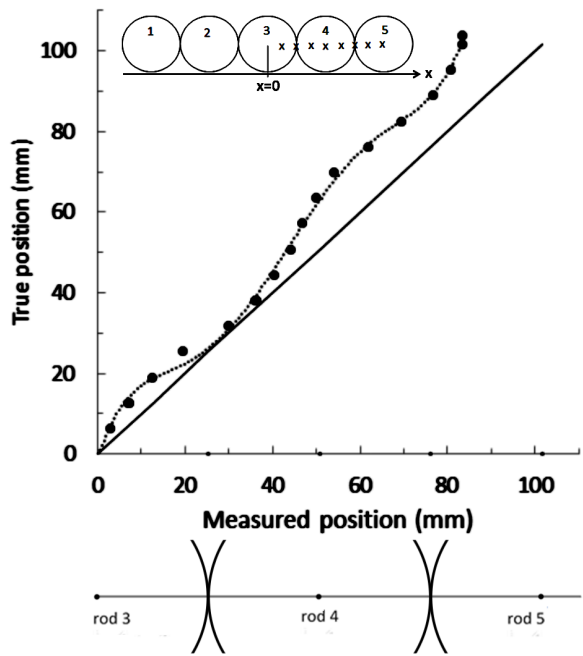

Figure 9. importance of corrections in GCI. The calculated position of the light source is compared to the known position for a 1D array of PMT rods. The source is shifted right starting from the center of the apparatus at $x = 0$. Several positions are shown in the inset. The solid line is the ideal case where the calculated position is equal to the location of the light source. The dotted line is a 5th degree polynomial fit of the true positions vs. the measured positions.

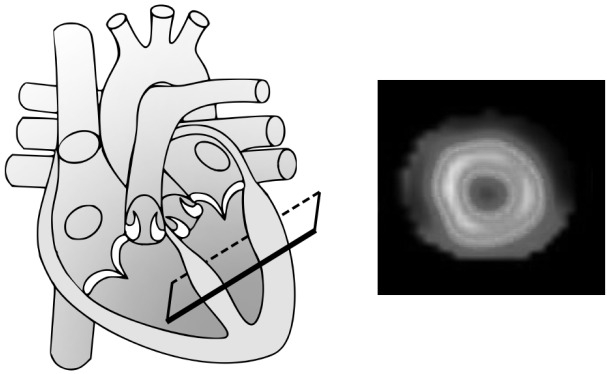

Figure 10. Cross-sectional image of the heart using SPECT. Image of the left ventricle wall was obtained using MPI, a technique that is conducted to determine areas of damaged coronary muscle. Image plane (left) and light gray regions (right) show where blood is flowing in the muscle. A full set of color image slices from three perpendicular directions appears in the supplementary material.[24]